\documentclass[twocolumn,prl,superscriptaddress,showpacs,preprintnumbers,amsmath,amssymb]{revtex4}

\usepackage[ansinew,latin1]{inputenc}
\usepackage{psfrag}
\usepackage{array}
\usepackage{amssymb}
\usepackage{amsmath}
\usepackage{graphicx}

\usepackage{color}
\newcommand{\red}{}

\begin{document}

\title{Polarized $^{3}$He as a probe for short range spin-dependent interactions}
\def\ILL{Institut Laue--Langevin, F--38042 Grenoble Cedex, France}
\def\ECTUM{Excellence Cluster `Universe', Technische Universit\"at M\"unchen, D--85748 Garching, Germany}

\author{A.~K.~Petukhov}     	\affiliation{\ILL}
\author{G. Pignol}          	\affiliation{\ECTUM}
\author{D. Jullien}          	\affiliation{\ILL}
\author{K. H. Andersen}         \affiliation{\ILL}

\date{\today}

\begin{abstract}
We have studied the relaxation of a spin-polarized gas in a magnetic field, in the presence of short-range spin-dependent interactions. 
As a main result we have established a link between the specific properties of the interaction and the dependence of the spin-relaxation rate on the magnitude of the holding magnetic field. 
This allows us to formulate a new, extremely sensitive method to study (pseudo-) magnetic properties at the submillimeter scale, which are difficult to access by other means. 
The method has been used as a probe for nucleon-nucleon axionlike P,T violating interactions which yields a two-order-of-magnitude improved constraint on the coupling strength ($g_s g_p$) as a function of the force range ($\lambda$): $g_s g_p \lambda^2 < 3 \times 10^{-27}$~m$^2$.
\end{abstract}


\maketitle

Hyperpolarized $^3$He is currently applied to a wide variety of scientific and medical problems. 
They include magnetic resonance imaging, spin-polarized targets, surface science, probing of biological systems, and precision measurements in fundamental physics \cite{Coulter,Batz,Gentile,Page,Andersen,Stewart,Keiderling,Hutanu,Lelievre-Berna,Chen}. 
One of the unique properties of polarized $^3$He is the very long spin-relaxation time constant which can be of the order of thousands of hours \cite{Rich,Parnell}, making polarized $^{3}$He extremely sensitive to any spin-dependent interaction. 
It is well known that the presence of a magnetic field gradient in a cell containing spin-polarized gas significantly affects the spin-relaxation. The origin of this relaxation mechanism is the loss of phase coherence of the atoms due to the fluctuating magnetic field seen by the atoms as they diffuse throughout the cell. 
Spin relaxation of a gas in the presence of a uniform gradient has been analyzed by Cates, Schaefer and Happer using a perturbation theory approach \cite{Cates} and by McGregor \cite{McGregor} within the framework of Redfield theory. 
In the present work we focus on spin-relaxation phenomena due to the field gradient decaying over distances much shorter than the cell size. 
This results in new expressions for the relaxation rates which agree with the already known result \cite{Cates,McGregor} in the limit of high pressure and high magnetic field (adiabatic regime) and in the uniform gradient limit  for very low pressure and magnetic field (``motion narrowing'' regime). 
A broad transitional regime of spin motion has been discovered which carries the complete information on the strength of this short-range field as well as on its spatial dependence. 
This finding allows us to propose a new very sensitive method to study short-range spin-dependent interactions from various origins, e.g. the magnetic field of a strongly diluted ferromagnetic sample or the pseudomagnetic field of hypothetical axionlike forces \cite{Weinberg,Wilezek,Moody}. 
To test the power of the method we have performed an experiment with polarized $^3$He measuring the longitudinal relaxation rate as a function of applied magnetic field. The experiment results in a new, stronger constraint on the axionlike interactions.

Suppose the magnetic field in a cell may be described by a homogeneous magnetic field ${\bf B}_0$ and weak inhomogeneous field ${\bf b}({\bf r})$: ${\bf B}({\bf r}) = {\bf B}_0 + {\bf b}({\bf r})$, with $<{\bf b}({\bf r})> = 0$. 
According to Slichter \cite{Slichter}, from the Redfield theory of spin-relaxation due to a randomly-fluctuating magnetic field, the spin-relaxation rates are given by 
\begin{eqnarray}
\label{Gamma1}
\Gamma_1 & = \frac{1}{T_1} = & \frac{\gamma^2}{2} \left[ S_x(\omega) + S_y(\omega) \right], \\ 
\label{Gamma2}
\Gamma_2 & = \frac{1}{T_2} = & \frac{\gamma^2}{4} \left[ S_x(\omega) + S_y(\omega) + 2 S_z(0) \right].
\end{eqnarray}
Here ${T_1}$ is the time constant for the longitudinal relaxation rate, ${T_2}$ is the time constant for the transversal
relaxation , $\gamma$ is the gyromagnetic ratio for the atoms of the gas ($\gamma \approx 2.04 \times 10^4 \ {\rm s}^{-1} \ {\rm G}^{-1}$ for $^3$He), and $\omega = \gamma B_0$ is the Larmor frequency. 
The functions $S_{k = x, y, z}(\omega)$ are the Fourier transform components of the magnetic field autocorrelation function $R_k(\tau) = \left< b_k(t) b_k(t + \tau)\right>$, where the ensemble average can be evaluated as
\begin{equation}
\label{autocorrelation}
R_k(t - t_0) = \iint \rho(r_0, t_0)\rho(r, t|r_0, t_0) b_k(r)b_k(r_0) d^3r d^3r_0, 
\end{equation}
knowing the conditional probability density $\rho(r, t|r_0, t_0)$ for an atom sitting at time $t_0$ at $r_0$ to be found at later time $t$ at a position $r$. The factor $\rho(r_0, t_0)$ is the single probability density.  
For times $|t-t_0|$ much longer than the mean time between atomic collisions, the conditional density obeys the diffusion equation \cite{Papoulis}, with the constraint that the initial $\rho(r,t_0|r_0,t_0) = \delta(r - r_0)$ and reflection boundary conditions. 
Let us consider a rectangular cell of length $L$ with square base of size $R$. 
Let $x$ be the cell axis so that the square ends occur at $x = \pm L/2$, a homogeneous field ${\bf B}_0$ directed along $z$ axis, and an inhomogeneous magnetic field ${\bf b}(x)$ directed along $x$ axis. 
For this geometry the problem becomes unidimensional with a known analytical solution for $\rho$ \cite{McGregor}. 
Substituting this solution in (\ref{autocorrelation}) and taking into account that in our case $\rho(r_0, t_0)=1/L$ we get the following expression for the autocorrelation: 
\begin{eqnarray}
& & R_x(\tau) = 2 \sum_{n=0}^{+\infty} e^{-\tau/\tau_n} b_{x, n}^2 \\
\nonumber
& & \tau_n = \frac{\tau_L}{\pi^2(2n+1)^2} \quad {\rm and} \\
& & b_{x, n} = \int_{-L/2}^{L/2} b(x) \sin((2n+1)\pi x/L) \frac{dx}{L}
\end{eqnarray}
where we have introduced the characteristic diffusion time constant $\tau_L = L^2/D$ , $D$ is the diffusion coefficient of the gas. 
Taking the Fourier transform gives 
\begin{equation}
\Gamma_1(\omega) = \frac{\gamma^2}{2} S_x (\omega) = 2 \gamma^2 \sum_{n = 0}^{+\infty} \frac{\tau_n}{1+(\omega \tau_n)^2} \ b_{x,n}^2. 
\end{equation}
To proceed further we have to specify the field $b(x)$. 
To take a concrete example, we consider the macroscopic pseudomagnetic field representing an axionlike interaction of polarized $^3$He with the nucleons in the cell walls \cite{Zimmer}:
\begin{eqnarray}
\label{axionlike}
& b(x) & = b_a \left(e^{-(L/2+x)/\lambda} - e^{-(L/2-x)/\lambda} \right), \\ 
& b_a & = \frac{\hbar \lambda}{2 \gamma m_n} N g_s g_p \left(1 - e^{-d/\lambda} \right).
\end{eqnarray}
Here, $x$ is the distancee from the wall, $g_s$ and $g_p$ are dimensionless scalar and pseudoscalar couplings between the nucleon and the axionlike particle, $\lambda = \frac{\hbar}{m_a c}$ is the force range, $m_n$ is the nucleon mass, $N$ is the nucleon number density and $d$ is the thickness of the walls. 
Substituting (\ref{axionlike}) in (5)-(6) we arrive at a general expression for the longitudinal relaxation on the one-dimensional case: 
\begin{widetext}
\begin{equation}
\label{BigFormula}
\Gamma_1 = (\gamma b_a)^2 \tau_{\lambda} \frac{(1+e^{-L/\lambda})^2}{\sqrt{\phi_L/2} (1+\phi_\lambda^2)^2} 
\begin{pmatrix}
\frac{(1 - \phi_\lambda(\phi_\lambda+2)) \sin \sqrt{\phi_L/2} + (1 - \phi_\lambda(\phi_\lambda-2)) \sinh \sqrt{\phi_L/2}}{\cos \sqrt{\phi_L/2} + \cosh \sqrt{\phi_L/2}} \\
 + \frac{1}{2} \sqrt{\phi_L/2} \ \left( {\rm sech}\frac{L}{2 \lambda} \right)^2 \ \left(1+\phi_\lambda^2 + \frac{\lambda}{L} (\phi_\lambda^2 - 3) \sinh \frac{L}{\lambda} \right)
\end{pmatrix}
\end{equation}
\end{widetext}
where $\phi_L = \omega \tau_L$, $\phi_\lambda = \omega \tau_\lambda$ and $\tau_\lambda=\lambda^2/D$. 

Our result (\ref{BigFormula}) is illustrated in Fig. 1. We can distinguish three regimes of relaxation: relaxation in a low magnetic field $\omega \ll \tau_L^{-1}$ with $\Gamma_1 {\red \propto} const$, relaxation in a moderate field $\tau_L^{-1} \ll \omega \ll \tau_{\lambda}^{-1}$ with $\Gamma_1 {\red \propto} \omega^{-1/2}$, and relaxation in a high field $\omega \gg \tau_\lambda^{-1}$ with $\Gamma_1 {\red \propto} \omega^{-2}$. 

\begin{figure}
\includegraphics[width=.78\linewidth]{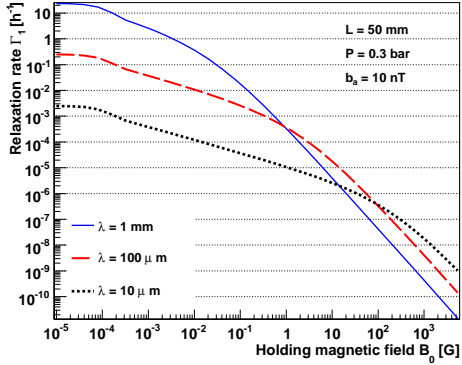}
\caption{
Relaxation rate due to a short-range gradient magnetic field (\ref{axionlike}) versus the magnitude of the homogeneous magnetic field calculated according to (\ref{BigFormula}) for three different values for the range $\lambda$. 
} \label{Gamma1Theo}
\end{figure}

The magnetic field value corresponding to the transition between the latter two regimes depends on the force range $\lambda$: the smaller $\lambda$ is, the higher magnetic field is needed. For an extremely sharp correlation function $R_k(\tau)$, like a Dirac delta-function, the Fourier transform $S_k(\omega)$ is constant and the corresponding relaxation rate is independent of magnetic field (relaxation due to dipole-dipole interaction in atomic collisions \cite{Newbery} or due to collisions with walls \cite{Deninger}). 
It can be shown that for the ``motion narrowing'' regime, $\omega \ll \tau_L^{-1}$:
\begin{eqnarray}
\label{Gamma1MN_large_lambda}
\Gamma_1(0) \approx & \frac{\gamma^2}{30} \frac{L^4}{D} \frac{b_a^2}{\lambda^2} \ & {\rm for} \quad \lambda \gg L, \\
\label{Gamma1MN_small_lambda}
\Gamma_1(0) \approx &  (\gamma b_a)^2 \tau_\lambda = \left< |\gamma b(x)| \right>^2 \tau_L \ & {\rm for} \quad \lambda \ll L. 
\end{eqnarray}

The expression (\ref{Gamma1MN_large_lambda}) agrees, with numerical factors of order one, with that \cite{Cates,McGregor} obtained for a spherical cell in a magnetic field with a uniform gradient. 
The expression (\ref{Gamma1MN_small_lambda}) 
shows that in the "motion narrowing" regime the relaxation is governed by 
{\red the mean squared} 
value of the inhomogeneous field, and by the time required for an atom
to diffuse over the cell. 
In the limit of very low pressure, when the 
mean-free path $l_{\rm coll}$ is larger than the cell size $L$, the natural time scale is $\tau_L =L/v$, $v$ being the particle velocity. 
{\red With} this replacement our expression (\ref{Gamma1MN_small_lambda}) agrees with \cite{Ignatovich} obtained for ultracold neutron depolarization. 

In the opposite ``adiabatic'' limit $\omega \gg \tau_\lambda^{-1} \gg \tau_L^{-1}$, our result (\ref{BigFormula}) reduces to
\begin{equation}
\Gamma_1 \approx D \frac{(\gamma b_a)^2}{\omega^2 \lambda L} \approx 2 D \left<\right( \frac{1}{B_0} \frac{d B_x}{dx} \left)^2 \right>
\end{equation}
also in agreement with \cite{Cates,McGregor}. 
For a finite-size rectangular cell and short-range $\lambda~\ll~R,L$ axionlike interactions with the cell walls, all components of the pseudomagnetic field $b_x(x)$, $b_y(y)$, and $b_z(z)$ need to be taken into account. 
Since the Brownian motions of the three coordinate directions are independent, from (\ref{Gamma1}) follows $\Gamma_1=\Gamma_{1,x}+\Gamma_{1,y}$, where both $\Gamma_{1,x}$ and $\Gamma_{1,y}$ are given by our expression (\ref{BigFormula}) for the one-dimensional geometry. 
The transverse relaxation rate follows from eqns. (\ref{Gamma2}) and reads, for a rectangular (or cubic) cell
\begin{equation}
\label{Gamma2_Cube}
\Gamma_2 = \Gamma_1(0) = 2 (\gamma b_a)^2 \tau_\lambda \quad {\rm for} \quad \lambda \ll R,L. 
\end{equation} 
We expect similar result for other geometries (cylindrical, or spherical) since (\ref{Gamma2_Cube}) is independent of the cell size. 
We checked by Monte Carlo simulation that this is indeed the case at the  $10 \%$ level of precision. 

We now apply our theoretical results to search for an exotic short-range axionlike interaction. 
Assuming for the time being no extra interactions, the experimental spin-relaxation is determined by contributions from three sources: $\Gamma_{1exp} = \Gamma_{dd} + \Gamma_{wall} + \Gamma_m$, where $\Gamma_{dd}$ is the dipole-dipole relaxation due to atomic collisions, $\Gamma_{wall}$ is due to the $^3$He spin relaxation on the walls of the cell and $\Gamma_m$ is due to magnetic field inhomogeneities. 
$\Gamma_{dd}$ and $\Gamma_{wall}$ are expected to be independent of magnetic field. 
If $B_0$ is high, so that $\omega \gg \tau_L^{-1}$, the relaxation due to magnetic field inhomogeneities can be decomposed into two components as follows:
\begin{equation}
\Gamma_m \approx D \left( \left< |g_i|^2 \right> + \left< |g_e|^2 \right> \right)/B_0^2 = \Gamma_{mi} + \Gamma_{me}
\end{equation}
where $g_i$ is the gradient due to inhomogeneities of the holding magnetic field $B_0$ and $g_e$ is the gradient caused by the external magnetic environment. 
Since $g_i {\red \propto} B_0$, the relaxation  $\Gamma_{me}$ due to the magnetic field gradient caused by the environment is the only term that depends on $B_0$. 
As all external magnetic sources are far away from the cell containing the polarized $^3$He, the relaxation is only affected by magnetic sources which are large compared to the cell size, and hence providing a nearly uniform gradient over the cell volume. 
Thus, we can expect that $\Gamma_{me}$ scales as $B_0^{-2}$. 
Finally, we can write:
\begin{equation}
\label{Gamma1Standard}
\Gamma_{1exp}(B_0) = \Gamma_{dd,wall,mi} + D |g_e|^2 \, B_0^{-2}.
\end{equation}
The relaxation due to short-range spin-dependent forces (\ref{BigFormula}) depends very differently on the magnetic field compared to the simple law (\ref{Gamma1Standard}), and can thus be separated from the other sources of relaxation. 
To take advantage of this feature we performed measurements of the longitudinal relaxation of $^3$He as a function of magnetic field using a cylindrical alumino-silicate glass cell (GE180, 5~cm diameter, 10~cm long, 3~mm wall thickness) filled with polarized $^3$He gas (75\% initial polarization at 0.3 bars pressure) placed inside a self-screening ``Magic box'' magneto-static cavity \cite{Petoukhov}. 
More details of the experiment will be published elsewhere. 
The experimental results obtained are shown in Fig.2 together with a fit of expression (\ref{Gamma1Standard}):
\begin{eqnarray}
\nonumber
\Gamma_{dd,wall,mi} & = & (9.71 \pm 0.16 ) \times 10^{-3} \ {\rm h}^{-1}, \\
|g_e| & = &  (2.70 \pm 0.05) \times 10^{-2} \ {\rm G/m}. 
\end{eqnarray}

\begin{figure}
\includegraphics[width=.78\linewidth]{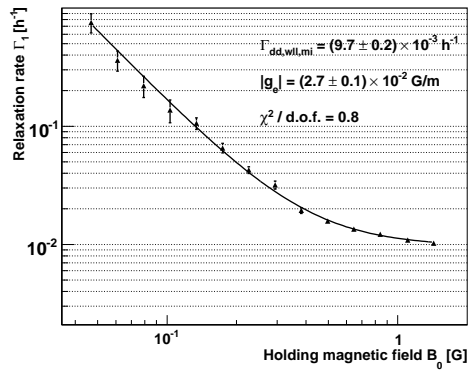}
\caption{
Experimentally measured longitudinal relaxation rate of polarized $^{3}$He versus magnitude of holding magnetic field (points). 
Solid line shows a fit of the theoretical prediction (15) for no short-range spin-dependent forces. 
} \label{Gamma1Exp}
\end{figure}

\begin{figure}
\includegraphics[width=0.87\linewidth]{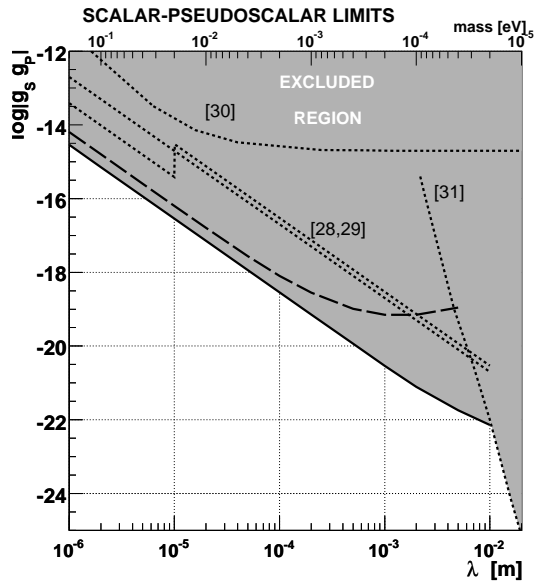}
\caption{
Constraints to the coupling constant product of axionlike particles to nucleons $g_s g_p$ as a function of the range $\lambda$ of the macroscopic interaction. 
Bold solid line: from long $T_2$ \cite{Gemmel}  (present work); 
bold dashed line: from scanning the magnetic field magnitude (present work); 
{\red dotted lines:} from UCN precession and depolarization 
\cite{SerebrovDepol,Serebrov}; 
from UCN gravitational levels \cite{Baessler}; 
from mercury spin precession \cite{Youdin}. 
} \label{Exclusion}
\end{figure}

Now we consider the addition of an axion rate (\ref{BigFormula}) to the normal expression (\ref{Gamma1Standard}). 
The fit involves now two additional free parameters $\lambda$ and $b_a$, 
it yields a result compatible with zero for $b_a$. 
Thus the experimental data shows no evidence for a new axionlike interaction. 
In order to set an upper limit on the strength of an axionlike interaction,  
the fit was performed again with various fixed values of $\lambda$ ranging from one micron to one centimeter. 
For each such value of $\lambda$ an upper limit on the product $g_s g_p$ was derived from the fit, which is shown as bold dashed line in Fig. \ref{Exclusion}. 
A previous attempt \cite{Pokotilovski} to determine constraints on axionlike forces in the range $10^{-6} \, {\rm m} < \lambda < 10^{-2} \, {\rm m}$ using available $^3$He relaxation data \cite{Rich,Parnell} appears to be flawed. 
That analysis was based on expression (12), which is only valid in the limit of high pressure and high magnetic field: $\omega \gg \tau_\lambda^{-1} \gg \tau_L^{-1}$, which corresponds to $\lambda > 10^{-4}$~m for the experimental conditions \cite{Rich,Parnell}. 

During the time that the present studies were performed, new experimental data on the very long time constant of the transverse relaxation in the ``motion narrowing'' regime became available from the shielded room BMSR-2 in Berlin \cite{Gemmel}. 
We have performed an analysis of these data within our theory. 
For the experimental conditions of \cite{Gemmel} the transversal relaxation time constant may be written as follows \cite{Cates,McGregor}:
\begin{equation}
\label{T2}
\frac{1}{T_2} = \frac{1}{T_1} + \frac{4 \gamma^2 R^4}{175 D} \left( \nabla B_x^2 + \nabla B_y^2 + 2 \nabla B_z^2 \right)
\end{equation}
where the first term represents the field-independent longitudinal relaxation due to interaction with walls and atomic collisions, while the second term represents the relaxation due to long-range gradients of the magnetic field. 
The first term {\red (\ref{T2})} was measured in a high magnetic field: $T_1 = 85 \pm 5 \ {\rm h}$.  
The second term was estimated to be $T_{2m} = 370 \pm 64 \ {\rm h}$ from the measured values of the gradients in {\red (\ref{T2})}. 
The authors \cite{Gemmel} compared the sum of these two terms $T_{2pred} = 69 \pm 4 \ {\rm h}$: with the experimentally measured $T_{2exp} = 60.1 \pm 0.1 \ {\rm h}$ and concluded that there was agreement between the theory and experiment. 
We can interpret the difference between these two relaxation time constants in terms of the effect due to axionlike interactions, via (8) and (13):
\begin{eqnarray}
\Gamma_{2a} = \frac{1}{T_{2exp}} - \frac{1}{T_{2pred}} = (2.1 \pm 0.8)\times 10^{-3} \ {\rm h}^{-1} \\
\label{NewLimit}
g_s g_p \lambda^2 < 3 \times 10^{-27} \ {\rm m}^{2}, \quad 95 \% \ {\rm C.L.}
\end{eqnarray}
A similar result was recently presented in \cite{Fu}.
Although the mean-free path between atomic collisions was as large as $l_c \approx 8\times 10^{-5}$~m, the limit (\ref{NewLimit}) still holds even for a new interaction in the micrometer range. 
Indeed, the diffusion regime approach, in particular Eq. (13), is valid as long as the mean-free path is much smaller than the cell size. 
This delicate point has been checked by Monte Carlo calculations based on Eq. (\ref{Gamma2}). 
The new upper limit (\ref{NewLimit}) is illustrated in Fig.\ref{Exclusion}  with the bold solid line. 
 
{\red 
Spin relaxation of $^3$He provides a sensitive method to study sub-millimeter magnetic properties or axionlike interactions. 
Analysing recent measurements \cite{Gemmel} performed in the best magnetically shielded room yields an upper limit on axionlike interactions almost 2 orders of magnitude better than that obtained from the published data on UCN spin precession \cite{Serebrov}. 
A new method is formulated, supported by a first experiment: scanning the relaxation rate $\Gamma_1$ on the holding magnetic field $B_0$. 
In general, this method samples the inhomogeneous field $b(x)$ at different spatial scales. 
Information about $b(x)$ can be extracted from comparison of the data $\Gamma_1(\omega)$ with the expectations of the model. 
In the particular case of the simple parametrization (\ref{axionlike}), one can extract the amplitude $b_a$ and the range parameter $\lambda$. 
Our first experiment could be dramatically improved if performed in a ``zero field room'' such as BMSR-2, where the external gradient of the magnetic field is 4 orders of magnitude lower than in our ``Magic box''. 
}

We are grateful to V. Nesvizhesky for attracting our attention to the axion problem and to E. Kats and R. Whitney for valuable discussions. 
One of us (A. P.) expresses his gratitude to C. Fu and T. Gentile for fruitful discussion during a visit at NIST. 


\end{document}